\documentstyle[12pt]{article}

\begin{document}

\begin{flushright}
hep-th/9510183 \\
preprint UT-728  \\
October, 1995\\
\end{flushright}

\bigskip

\begin{center}
{\large \bf Prepotentials of $N=2$ Supersymmetric Gauge Theories and
Soliton Equations}
\end{center}

\bigskip

\begin{center}
Tohru Eguchi

\medskip

{\it Department of Physics

\medskip

Faculty of Science

\medskip

University of Tokyo

\medskip

Tokyo 113, Japan}

\bigskip

and

\bigskip

Sung-Kil Yang

\medskip

{\it Institute of Physics

\medskip

University of Tsukuba

\medskip

Ibaraki 305, Japan}
\end{center}

\bigskip

\bigskip

\begin{abstract}
Using recently proposed soliton equations we derive a basic identity for the
scaling violation of $N=2$ supersymmetric gauge theories
$\sum_i a_i\partial F/\partial a_i-2F=8 \pi i b_1 u$. Here $F$ is the
prepotential, $a_i$'s are
the expectation values of the scalar fields in the vector multiplet,
$u=1/2\, {\rm Tr}\, \langle\phi^2\rangle$ and $b_1$
is the coefficient of the one-loop $\beta$-function.
This equation holds in the Coulomb
branch of all $N=2$ supersymmetric gauge theories coupled with massless
matter.
\end{abstract}

\newpage

Recently there have been some major progress in our understanding of the
non-perturbative
strong coupling behavior of 4 dimensional supersymmetric gauge theories and
string theories in various dimensions based on the idea of
S (strong-weak) duality
\cite{Sen}-\cite{Sch}.
In the case of $N=2$ supersymmetric gauge fields for various gauge groups,
many of their low energy effective
Lagrangians have been determined exactly with or without the presence of
matter fields \cite{AF}-\cite{H}.
There now exist a considerable amount of data for the algebraic
curves and differential forms which are used to compute the prepotentials of
the $N=2$ effective Lagrangian.

Quite recently an idea has been proposed which may possibly organize these data
within the framework of some known integrable systems \cite{GKMMM}-\cite{NT}.
In this article we will adopt a machinery from the soliton theory and derive
an important relation obeyed by the prepotentials $F$ of general $N=2$
supersymmetric
gauge theories coupled to massless matter fields

\begin{equation}
\sum_{i=1}^{r} a_i {\partial F \over \partial a_i} - 2F =8\pi i\, b_1 u .
\label{eq:1}
\end{equation}
Here $b_1$ is the coefficient of the one-loop $\beta$-function and
$a_i \ \ (i=1,\cdots,r)$
are the expectation values of the Cartan components of the scalar fields
in the vector multiplet.
$r$ is the rank of the gauge group $G$ and
$u=1/2\, {\rm Tr}\, \langle\phi^2\rangle$.
Eq.(\ref{eq:1}) holds in the Coulomb branch of the theory.
The right-hand-side of the above equation shows precisely the amount of the
scaling violation dictated by the $\beta$-function.

The simplest version of the formula (\ref{eq:1}) was obtained previously by
Matone for the case $G=SU(2), N_f=0$ making use of the Picard-Fuchs (P-F)
equation \cite{M}.
In the case of the $SU(2)$ theory it is easy to generalize the
analysis to the
case of matter fields $N_f \neq 0$. The method of P-F equations, however,
becomes messy and complicated when we go to higher rank groups and we need an
alternative method of derivation. As we shall see,
the machinery of soliton equation is particularly suited to our purpose and
it is easy to derive the formula (\ref{eq:1}).

During the preparation of this article there appeared a new preprint
``On the Relation
Between the Holomorphic Prepotentials and the Quantum Moduli in SUSY
Gauge Theories'' (hep-th/9510129) by J. Sonnenschein, S. Theisen and
S. Yankielowicz \cite{STY}
which proves (\ref{eq:1}) from a somewhat different starting point.

Let us first recall the $SU(2)$ case without matter fields. The curve is given
by $y^2=(x^2-\Lambda^4)(x-u)$\cite{SWa} and the period integral of
the meromorphic
1-form $\lambda={\sqrt{2}\over 2\pi}{(x-u) \over y}dx$
obeys the P-F equation \cite{KLT,CDF}
\begin{equation}
\frac{d^2 \omega(u)}{du^2}+ \frac{1}{4(u^2-\Lambda^4)}\omega(u)=0.
\label{eq:2}
\end{equation}
Since eq.(\ref{eq:2}) has no first-derivative term,
the Wronskian made out of its solutions $a(u),a_D(u)$ is $u$-independent
\begin{equation}
a(u)a_D(u)' - a_D(u)a(u)'= \mbox{constant}.
\label{eq:3}
\end{equation}
(Note that the Wronskian is invariant under the $SL(2,{\it R})$ transformation
among
$a$ and $a_D$.)
By integrating (\ref{eq:3}) over $u$ we find
\begin{equation}
a(u)\frac{\partial F}{\partial a(u)} -2 F(u) = \mbox{constant}\times u.
\label{eq:4}
\end{equation}
Recall that $a_D=\partial F/\partial a$.
Evaluating both sides of (\ref{eq:4}) at
$u=\infty$ by making use of the weak-coupling expansion
\begin{eqnarray}
&&F\approx\frac{i}{2 \pi} a^2 \big(\log a^2/\Lambda^2 +O((\Lambda/a)^4)\big),
\label{}
\\
&&u \approx \frac{1}{2} a^2 (1+O((\Lambda/a)^4)),
\label{}
\end{eqnarray}
we find the value of the constant=$4i/2 \pi$. In fact this value
is proportional to
the coefficient of the $a^2\log a^2$ term in $F$ and hence
the one-loop $\beta$ function.

It is straightforward to
generalize the above computation to the $N_f\neq 0$ cases ($N_f=1,2,3$).
We find that the first-derivative term is always absent
in the P-F equation \cite{IY} and hence the
Wronskian is a constant proportional to $b_1=(4-N_f)/16 \pi^2$.

We recall that in $N=2$ supersymmetric Yang-Mills theories
with the general
gauge group, one introduces a hyperelliptic curve $\Sigma$ and a meromorphic
differential $\lambda$ whose
periods give expectation values
$a_i, a_i^D  (i=1,\cdots,r)$ \cite{AF,KLTY}.
The meromorphic 1-form $\lambda$ has a double pole
at $\infty$ whose residue is proportional to $b_1$. When one couples massive
matter, $\lambda$ may also have a simple pole at $\infty$ whose residue is a
sum of the masses $m_i$ with half-integer coefficients. P-F equations in the
case of higher rank groups are partial differential equations and their
analysis seems to involve a number of technical complications.
In the following we, instead, use an
approach based on soliton theory, i.e. Whitham's method of adiabatic
perturbation of integrable systems \cite{Whitham}.

Let us now describe in some detail the structure of the Whitham dynamics
which has been proposed to be relevant for the analysis of $N=2$ supersymmetric
gauge theories \cite{GKMMM}-\cite{NT}.
We follow the presentation of ref.\cite{NT}.
In \cite{GKMMM,NT} new ``time'' variables
$T_n\ (n=0,1,\cdots)$ are introduced which are coupled
to the $(n+1)$-th order pole of the differential $\lambda$.
It is assumed that $\lambda$ satisfies the following equations
\begin{equation}
\frac{\partial \lambda}{\partial a_i}=\omega_i \ (i=1,\cdots,r), \ \
\frac{\partial \lambda}{\partial T_0}=\Omega_0, \ \
\frac{\partial \lambda}{\partial T_n}=\Omega_n  \ (n=1,2,\cdots).
\label{eq:7}
\end{equation}
Here $\omega_i$ are holomorphic differentials normalized as $\oint_{\alpha_j}
 \omega_i=\delta_{ij}$. $\Omega_n,\Omega_0$ are meromorphic differentials of
the 2nd and 3rd kind, respectively, with a behavior
$\Omega_n\approx -n z^{-n-1}, \ \ \Omega_0\approx z^{-1}$
at $x=1/z=\infty$. $\Omega_n$ and $\Omega_0$ have vanishing periods for the
$\alpha$-cycles.
($\Omega_0$ behaves as
$\Omega_0\approx -z_*^{-1}$ at $x=1/z_*=\infty$ on the other Riemann sheet of
the hyperelliptic surface.)
Eq.(\ref{eq:7}) leads to the integrability conditions for the differentials
\begin{equation}
\frac{\partial \omega_j}{\partial a_i}=\frac{\partial \omega_i}{\partial a_j},
\ \ \frac{\partial \omega_i}{\partial T_n}
=\frac{\partial \Omega_n}{\partial a_i}, \ \
\frac{\partial \Omega_m}{\partial T_n}=\frac{\partial \Omega_n}{\partial T_m}.
\label{eq:8}
\end{equation}
The prepotential $F$ is defined by the equations
\begin{equation}
\frac{\partial F}{\partial a_i}=\oint_{\beta_i}\lambda, \ \ \
\frac{\partial F}{\partial T_n}=-2 \pi i\
\mbox{res}\Big(z^{-n}\lambda\Big), \ \ \
\frac{\partial F}{\partial T_0}=-2 \pi i\, \int _{z_*=0}^{z=0} \lambda ,
\label{eq:9}
\end{equation}
where ``res'' means a residue at $z=0$.
$\lambda$ then behaves at $\infty$  as
\begin{equation}
\lambda=\Big(-\sum_{n\ge 1} n T_n z^{-n-1}
+T_0 z^{-1}-\frac{1}{2\pi i}\sum _{n\ge 1} \frac{\partial F}
{\partial T_n} z^{n-1}\Big) dz.
\label{asym}
\end{equation}
Making use of the Riemann bilinear relations it is possible to check
integrability conditions (\ref{eq:8}). For instance
\begin{eqnarray}
&& \frac{\partial}{\partial T_n}\frac{\partial F}{\partial a_i}=\frac
{\partial}{\partial T_n}\oint_{\beta_i}\lambda
=\oint_{\beta_i}\Omega_n  \nonumber \\
&& \ \ \ \  =-\sum_{j=1}^r \Big(\oint_{\alpha_j}\Omega_n\oint_{\beta_j}\omega_i
-\oint_{\alpha_j}\omega_i\oint_{\beta_j}\Omega_n \Big)  \nonumber  \\
&& \ \ \ \ =-2 \pi i\   \mbox{res}\Big(\phi_n \omega_i\Big)=-2\pi i \
\mbox{res}\Big(z^{-n}\omega_i\Big), \\
&&\frac{\partial}{\partial a_i}\frac{\partial F}{\partial T_n}=-2 \pi i
\frac{\partial}{\partial a_i}\mbox{res} \Big(z^{-n}\lambda\Big)=-2 \pi i
\  \mbox{res} \Big(z^{-n}\omega_i\Big),
\label{}
\end{eqnarray}
where $\phi_n(z)=\int^{z}\Omega_n$.

Solutions of the Whitham dynamics eqs.(\ref{eq:7})-(\ref{asym})
become relevant to
the $N=2$ theory when they obey the homogeneity condition
\begin{equation}
\sum a_i \frac{\partial F}{\partial a_i}+T_0\frac{\partial F}{\partial T_0}
+\sum T_n\frac{\partial F}{\partial T_n}=2 F.
\label{hom}
\end{equation}
Eq.(\ref{asym}) together with
(\ref{hom}) leads to the relation \cite{NT}
\begin{equation}
\lambda= \sum_{i=1}^r a_i \omega_i + T_0 \Omega_0+\sum_{n=1}T_n\Omega_n .
\label{}
\end{equation}
The above equation has the form of the meromorphic differentials of $N=2$
theories when
$T_n$'s are put to zero except $T_0, T_1$.
Therefore, using the soliton theory we obtain an identity for
the prepotential in $N=2$ supersymmetric gauge
theories
\begin{equation}
\sum a_i \frac{\partial F}{\partial a_i}-2F=-T_0\frac{\partial F}{\partial T_0}
-T_1\frac{\partial F}{\partial T_1}.
\label{identity}
\end{equation}
Evaluation of $T_0,T_1,\partial F/\partial T_0$ is straightforward: we simply
read them off from the expansion coefficients of $\lambda$ at $\infty$
in (\ref{asym}).
On the other hand, $\partial F/\partial T_0$ involves a line integral of
the differential
$2 \pi i \int_{z_*}^z \lambda$
which is hard to evaluate in general. This problem is avoided if we consider
theories with massless matter fields where the parameter $T_0$ vanishes.
Therefore in supersymmetric Yang-Mills theories coupled to
massless hypermultiplets we obtain a basic formula
\begin{equation}
\sum a_i \frac{\partial F}{\partial a_i}-2F=
-T_1\frac{\partial F}{\partial T_1}=-2\pi i\, \mbox{res}(z\lambda)\, \mbox{res}
(z^{-1}\lambda).
\label{basic}
\end{equation}
We shall show that the right-hand-side of eq.(\ref{basic})
has the universal form
\begin{equation}
8\pi i\, b_1 u
\label{univ}
\end{equation}
in all the examples discussed below.
Note that in $SO(N_c)$ gauge theories coupled to the vector matter $T_0$
vanishes
due to symmetry $x\rightarrow -x$ of the curve.

We now compute the right-hand-side of eq.(\ref{basic}) in $N=2$
theories where curves and differentials are known explicitly.

\bigskip
{\it $SU(N_c)$ theory with matter in the vector representation} \cite{HO}

\bigskip

(a) $N_f<N_c$
\begin{eqnarray}
&&y^2=C(x)^2-\Lambda_{N_f}^{2N_c-N_f}G(x), \\
&&C(x)=x^{N_c}-\sum_{i=2}^{N_c}u_i x^{N_c-i},
\ \ G(x)=\prod_{i=1}^{N_f}(x+m_i),
\ \ u=u_2
\\
&&\lambda=\frac{x dx}{2\pi i y}\big(\frac{C G'}{2 G}-C'\big),
\ \ \ \ z=1/x
\nonumber \\
&&\ \ \ \approx \frac{dz}{2\pi i}\Big((\frac{N_f}{2}-N_c)z^{-2}-\frac{1}{2}
\sum_{i=1}^{N_f} m_i z^{-1}+(-2u+\frac{1}{2}\sum_{i=1}^{N_f} m_i^2)\Big).
\end{eqnarray}
Thus
\begin{equation}
T_1=\frac{-N_f/2+N_c}{2 \pi i},
\ \ T_0=-\frac{1}{4 \pi i}\sum_{i=1}^{N_f} m_i,
\ \ \frac{\partial F}{\partial T_1}=2u-\frac{1}{2}\sum_{i=1}^{N_f} m_i^2
\end{equation}
and we reproduce (\ref{eq:1}) in the massless
limit ($b_1=(2 N_c-N_f)/16\pi^2$).

\bigskip

(b) $N_f\geq N_c$
\begin{eqnarray}
&&y^2=F (x)^2-\Lambda_{N_f}^{2N_c-N_f}G(x),  \\
&&F(x)=C(x)+\frac{\Lambda_{N_f}^{2N_c-N_f}}{4}\sum_{i=0}^{N_f-N_c}
x^{N_f-N_c-i}t_i(m), \\
&&t_k(m)=\sum_{i_1<\cdots<i_k}m_{i_1}\cdots m_{i_k}.
\end{eqnarray}
We again find eq.(\ref{eq:1}) in the massless limit.

\bigskip

{\it $SO(N_c)$ gauge theory with matter in the vector representation} \cite{H}

\bigskip

(a) $N_f< N_c/2-2$ for $N_c$ even,
\ \ $N_f< N_c/2 -3/2$ for $N_c$ odd

\medskip

\begin{eqnarray}
&&y^2=C(x)^2-\Lambda_{N_f}^{2(N_c-2-N_f)}G(x), \\
&&C(x)=
\left\{
\begin{array}{ll}
\displaystyle{
\prod_{i=1}^{N_c/2}(x^2-e_i^2)=x^{N_c}-ux^{N_c-2}-\cdots ,}&
              \ N_c\ \mbox{even} \\
\displaystyle{
\prod_{i=1}^{(N_c-1)/2}(x^2-b_i^2)=x^{N_c-1}-ux^{N_c-3}-\cdots ,}&
              \ N_c\ \mbox{odd} \\
\end{array}\right.  \\
&&G(x)=x^d\prod_{i=1}^{N_f}(x^2-m_i^2), \quad
\left\{
\begin{array}{rl}   d=4& \mbox{for}\ N_c\ \mbox{even} \\
                    d=2& \mbox{for}\ N_c\ \mbox{odd}  \\
\end{array}\right.  \\
&&\lambda=\frac{x dx}{2 \pi iy}\Big(\frac{C G'}{2G}-C'\Big) .
\end{eqnarray}

\bigskip

(b) $N_f\geq N_c/2-2$ for $N_c$ even,
\ \ $N_f\geq N_c/2-3/2$ for $N_c$ odd

\medskip

\begin{eqnarray}
&&y^2=\big(C(x)+\Lambda^{2(N_c-N_f-2)}P(x)\big)^2-\Lambda_{N_f}^{2(N_c-2-N_f)}
G(x), \\
&&\lambda=\frac{x dx}{2 \pi iy}\Big(\frac{C G'}{2G}-C'\Big) .
\end
{eqnarray}
Here $P(x)$ is a polynomial of order $2N_f-(N_c-2)$ in $x$ and $m_i$.
We note that in $SO(N_c)$ gauge theories $2r$ periods
out of $2(2r-1)$ periods of
$\Sigma$ are independent due to $x\rightarrow -x$ symmetry of the curve.
In both cases (a), (b) we find
\begin{equation}
T_0=0, \ \ T_1=\frac{-N_f+N_c-2}{2\pi i}, \ \ \frac{\partial F}{\partial T_1}
= 2u-\sum_{i=1}^{N_f} m_i^2.
\label{}
\end{equation}
Thus eq.(\ref{eq:1}) holds also in $SO(N_c)$ theories
($b_1=2(N_c-2-N_f)/16 \pi^2$).

So far we have derived eq.(\ref{eq:1}) only in the case of $SU(N_c)$ and
$SO(N_c)$ gauge
theories with matter in the vector representations for which data on algebraic
curves
and differential forms are available. However, it seems natural
to conjecture that it holds in the Coulomb branch of all $N=2$ Yang-Mills
theories
with arbitrary gauge groups and arbitrary massless hypermultiplet
representations.
In order to discuss general cases, deeper understanding of the choice
of cycles and sub-spaces of Jacobians of $\Sigma$ seems necessary \cite{MW}.

Equation (\ref{eq:1}) will play a basic role
when 4 dimensional gauge theories are
embedded into supergravity and superstring theories.
In locally supersymmetric theories the variable $T_1$ will be
identified as the expectation value of the dilaton field as in the important
examples
worked out in ref.\cite{KKLMV}. {$T_n$} variables in eq.(\ref{hom})
restore the homogeneity of the prepotential destroyed by the non-vanishing
$\beta$-function and bring eq.(\ref{hom}) into a form known in the special
geometry of $N=2$ supergravity.
It will be very interesting to see if it is possible to
provide physical interpretation of the $T_n$ variables which may describe
the gravity sector of $N=2$ supergravity theories. Eq.(\ref{hom}) is
also reminiscent
of the Virasoro $L_0$ condition in 2 dimensional topological $\sigma$
model coupled to topological gravity \cite{Hori}.
It is interesting to see if there are analogues of $W$ conditions
in the case of higher rank $N=2$ gauge theories.

\vskip10mm
We would like to thank Dr. T. Nakatsu for discussions on his work. Research of
T.E. and S.K.Y. is supported in part by the Grant-in-Aid for Scientific
Research on Priority Area 213 ``Infinite Analysis'', Japan Ministry of
Education.

\end{document}